# Molecular dynamics simulations of a single stranded (ss) DNA

Subhasish Chatterjee[1,†], Bonnie Gersten[1], Siddarth Thakur[2], Alexander Burin[2]

[1]Department of Chemistry, Queens College and the Graduate Center of CUNY,
City University of New York, Flushing, NY 11367, USA
[2]Department of Chemistry, Tulane University, New Orleans, LA, 70118, USA

**ABSTRACT**

The objective of the present study was to develop an understanding of short single-stranded DNA (ssDNA) to assist the development of new DNA-based biosensors. A ssDNA model containing twelve bases was constructed from the 130-145 codon sequence of the p53 gene. Various thermodynamic macroscopic observables such as temperature, energy distributions, as well as root mean square deviation (RMSD) of the nucleic acid backbone of the ssDNA were studied using molecular dynamics (MD) simulations. The AMBER program was used for building the structural model of the ssDNA, and atomistic MD simulations in three different ensembles were carried out using the NAMD program. The microcanonical (NVE), conical (NVT) and isobaric–isothermal (NPT) ensembles were employed to compare the equilibrium characteristics of ssDNA in aqueous solutions. Our results indicate that the conformational stability of the ssDNA is dependent on the thermodynamic conditions.

[†] schatterjee@gc.cuny.edu

**Introduction**

In recent years, the specific and selective hybridization of nucleic acids have been utilized widely for the development of novel nanostructures, including nanoelectronics, nanomechanics and biosensing devices [1-2]. The specific molecular recognition property and hybridization phenomenon of DNA significantly control the sensitivity and selectivity of DNA based biosensors [2-4]. DNA possesses a polyanionic backbone, composed of alternating sugar and phosphate groups, and has four different bases, namely, Adenine (A), Guanine (G), Cytosine (C) and Thymine (T) [3]. The specific molecular recognition property of DNA originates from the selective base pairing: A binds to T and C binds to G [3, 4]. Apart from this distinctly characteristic base-pairing property, the polyanionic backbone controls the physiochemical properties of DNA such as flexibility, electrostatic properties, and binding capacity to cationic nanoparticulates [5, 6]. DNA-based biosensors fundamentally rely on the hybridization phenomenon, in which a single strand (ss) DNA binds selectively to its complementary strand under ambient conditions [2, 7]. As a result of the exposed bases and the lack of a comparatively rigid double helix structure, ssDNA shows considerable differences in electrostatic properties and flexibility compared with double strand (ds) DNA [3, 7]. Consequently, the ssDNA plays a significant role in the hybridization process as well as in the proper functioning of these nanostructure devices [2, 3, 7].

In view of the fact that the biosensing devices primarily use short DNA strands, this study focused on the 130-145 codon (15 codon) sequence of the p53 gene, which is involved in tumor suppression [8], and a mutation in this codon sequence of p53 can cause cancer, in particular, lung cancer. The specific objective of this study was to theoretically understand the conformational characteristics of short single strand of nucleic acids under various thermodynamic conditions. Accordingly, an understanding of the behavior of short strands of the DNA system can be employed to improve the hybridization process, associated with the characteristic role of the biosensing devices. Molecular dynamics (MD) simulations of the ssDNA were performed for this investigation.

**Simulation Method**

MD simulations can delineate the atomistic details of the DNA system because they compute atomic trajectories by solving equations of motion using empirical force fields that describe the actual atomic force in biomolecular systems [9]. The simulation was carried out by the NAMD molecular dynamics program [10], which uses a potential energy function that considers various bonding and nonbonding energy contributions, including bond stretching, bending, and torsional bonded interactions [10]. The force field is the mathematical description of the potential that atoms in the system experience. In this study, the AMBER force field was utilized, as it is widely applied to describe nucleic acid system [11]. The first, second, and third terms in the function describe bond stretching, bending, and torsional energy, respectively. The fourth term depicts non-bonded interaction; the fifth term describes electrostatic energy [11]. In the force field, chemical bonds are represented by harmonic potentials, which assume that bonds cannot be ruptured. This description of the bonds is only realistic close to the equilibrium distance. The bond force constants, estimated from spectroscopic methods or quantum mechanical calculations, determine the flexibility of a bond. Similarly, a harmonic potential describes the energy related to the alterations of bond angles, given by the second term in the AMBER force field. The rotational bond energies are depicted by the torsion angle potential functions, modeled by a periodic function and sum of cosine terms of angles. The potential functions representing nonbonded interactions are comprised of van der Waals and electrostatic forces. The long-range electrostatic interaction is a very important and challenging issue to address in obtaining a valid result in a biomolecular simulation [10, 12]. Since the Ewald method is very reliable for estimating electrostatic interactions in a spatially limited system, the particle–mesh Ewald (PME) method was adopted for a faster numerical computation of the electrostatic interaction in this study [10].

The MD simulation was carried out in the presence of explicit solvent molecules and $Na^+$ counter ions. The periodic boundary condition was employed to perform the simulation. A TIP3P water model in a simulation box was used for solvating the DNA system [11]. The structural parameters and coordinates necessary for the ssDNA system were obtained by taking into consideration a helix from double stranded B-DNA designed by the AMBER program with the above-mentioned p53 sequence (130-140 codon sequence) [11]. A single DNA strand containing 12 bases was prepared for this molecular dynamics study.

Energy minimization of the ssDNA and equilibration of the system at particular thermodynamic conditions are the two initial important steps to undertake a valid molecular dynamics study [9, 12]. The potential energy of the solvated ssDNA was minimized for 2000 steps to get the energy-minimized structure of ssDNA. The energy-minimization was done by application of the conjugate gradient method in the NAMD program [10]. Thereafter, the energy-minimized structure of the ssDNA was equilibrated while maintaining proper thermodynamic conditions of the ensemble of interest. The equilibration was performed for microcanonical (NVE), canonical (NVT), and isobaric-isothermal (NPT) ensembles. The temperature was maintained at 300 K for all ensembles. The pressure was maintained at 101325 Pa (1 atm) in the NPT ensemble. The consistency of temperature was maintained by performing Langevin Dynamics [9, 10]. The constant pressure control was executed by the Langevin piston Nose-Hoover method, available in the NAMD program [10]. VMD, a molecular viewing program, was used to visualize the dynamics of the system [13]. The achievement of the dynamic equilibrium of the system was evaluated by how well energy, pressure, and temperature (thermodynamic properties of the system) were distributed in the system over a certain amount of time [9, 12].

**Results and Discussion**

Thermodynamic observables (pressure, temperature, and volume) play an important role in setting up the proper ensemble for the simulation study [10, 14]. The distribution of energy (kinetic and potential energy) indicates the establishment of dynamic thermodynamic equilibrium. Since the force field in the MD simulations assumes that the bonded interactions (bonds, angles and dihedrals) maintain the characteristics of harmonic oscillators [9, 10], fluctuations in kinetic energy, and hence the temperature distribution can be attributed to the equilibrium of the ssDNA system. The simulation was carried out in the microcanonical ensemble (constant N (number of atoms), V (volume), and E (energy). The dynamics of system was observed for 100 ps after initial equilibration of 50 ps. The initial equilibration was performed at the constant temperature of 300 K.

The fluctuation in temperature during the dynamics study maintained the Gaussian distribution with a mean temperature of 303.99 K (**Figure 1**). The fluctuations in the temperature describe the effect of the finiteness of the system.

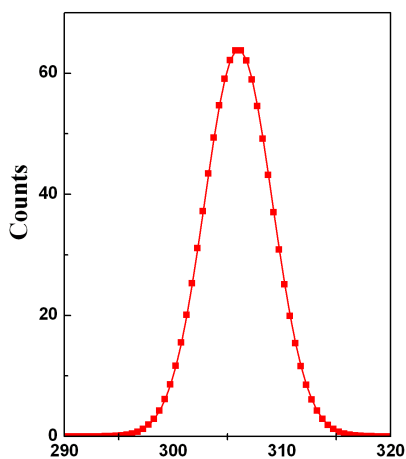

Figure 1: Temperature fluctuation observed in NVE ensemble of ssDNA

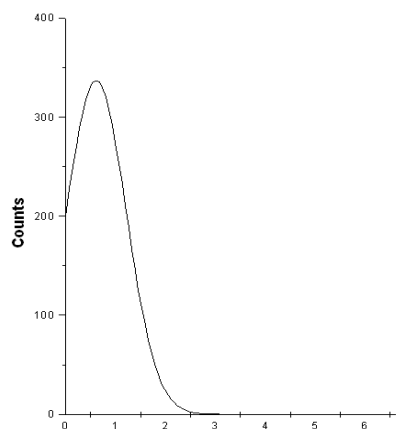

Figure 2: The distribution of kinetic energy of ssDNA system in NVT ensemble.

As the system reaches dynamic equilibrium, the total energy of the system approaches a constant value. The fluctuations in the total energy of the system depend on the number of atoms or particles present in the simulating system. The average kinetic energy resulted in the temperature of the solvated ssDNA system. The distribution of the kinetic energy of the ssDNA system confirms the establishment of physiological temperature. The kinetic energy distribution of solvated ssDNA system was computed by running the dynamics for 100 ps in the canonical (NVT) ensemble. The distribution of kinetic energy of the ssDNA system followed the Maxwell-Boltzmann distribution with the standard deviation of 0.602 Kcal/mole, which corresponded to a temperature of approximately 300 K (**Figure 2**). Thus, the desired equilibrium state was achieved by performing the proper sampling of the ssDNA configurations.

The Root Mean Square Deviation (RMSD) of the ssDNA backbone was computed for NVT and NPT ensembles (**Figures 3 and 4**). The RMSD provides the numerical measure of the difference between two structural states of the ssDNA [15, 16]. The conformational stability of the ssDNA plays a significant role in maintaining a proper equilibrium state. The RMSD of the nucleic acid backbone shows the conformational state of the ssDNA with the progress of the dynamics [15, 16]. As the RMSD of the nucleic acid strand changes over time, it provides an indication of the stability of the ssDNA at the specific state of equilibrium. For the NVT ensemble study, at first the ssDNA system was equilibrated for 50 ps at 300 K. The achievement of the dynamic equilibrium state was evaluated by examining the consistency in the total energy, temperature and volume. Thereafter, the dynamics of the system was studied for 50 ps. The RMSD of the nucleic acid backbone was calculated during this dynamics study to investigate the conformational change of the ssDNA from its apparent equilibrium structure. The RMSD of the ssDNA exhibited an approximately flattening curve (**Figure 3**) with the deviation of with the deviation of approximately 1.4 Å to 1.6 Å. This result shows a stable conformational state of the ssDNA under the prevailing thermodynamic conditions.

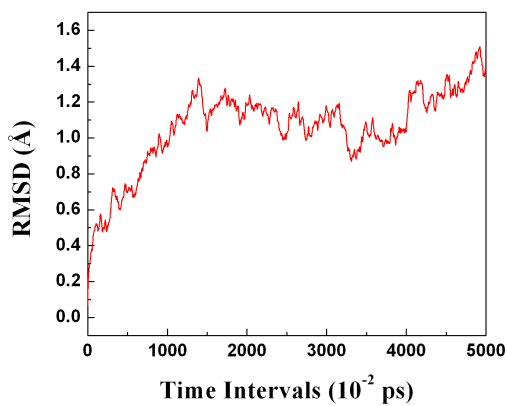

**Figure 3: RMSD of ssDNA backbone in NVT ensemble.**

The RMSD of the nucleic acid backbone was computed by a similar method for the NPT ensemble. The simulation was performed at 300 K temperature and 1 atm pressure. The ssDNA system was equilibrated at first for 50 ps, followed by the dynamics study of 400 ps. The RMSD showed a change of approximately 3Å with the progress of time (**Figure 4**) during the dynamics study. These results indicated that the ssDNA might have been undergoing structural change from its previous conformation over time, in spite of the fact that the system maintained the consistency in the equilibrium thermodynamic conditions.

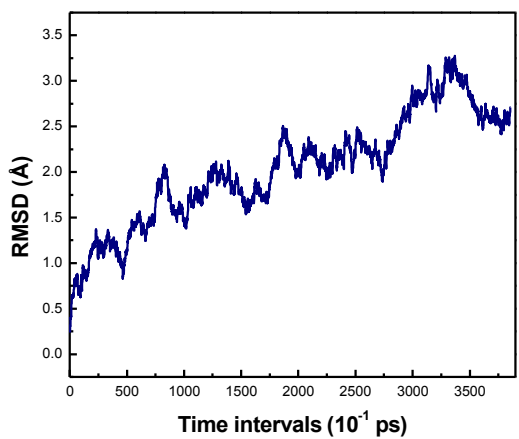

Figure 4: RMSD of ssDNA backbone in NPT ensemble

As the NPT ensemble resembles a closer description of the biological conditions, this latter study provides an idea of the conformational flexibility of the ssDNA under the ambient conditions. The proper thermodynamic conditions and the suitable conformational states of the ssDNA regulate the rate of the hybridization process. The selectivity and specificity of the hybridization of nucleic acids depend significantly on the conformational stability and flexibility of the ssDNA. Our study indicated the effect of thermodynamic conditions on the conformational state of the ssDNA, which could potentially modulate the qualitative features of DNA-based biosensing devices.

**Conclusions**

The conformational stability of the ssDNA depends on the thermodynamic conditions, and the flexibility of ssDNA undergoes alteration with the change in the thermodynamic parameters of the system.

*Acknowledgements*

The authors thank Prof. T. Bishop of the Center for Bioenvironmental Research at Tulane University, for his helpful suggestions. The authors thank Dr. Scott Grayson for his helpful comments. The work was supported in part by a grant from The City University of New York PSC-CUNY Research Award program and in part by the James D Watson NYSTAR Award.